\documentclass[aps,physrev,twocolumn,superscriptaddress,nofootinbib,balancelastpage]{revtex4-2} 
\usepackage[latin1]{inputenc}
\usepackage{amsmath,amssymb}
\usepackage{mathrsfs} 
\usepackage{siunitx}
\usepackage{braket}
\usepackage{calc}
\usepackage{esint}
\usepackage{tabularx}
\usepackage{dsfont}
\usepackage{color}
\usepackage{ifthen}
\usepackage[normalem]{ulem}% Fuer Durchstreichungen
\usepackage{graphicx}
\usepackage{seqsplit}
\usepackage[colorlinks,allcolors=black]{hyperref} 
\usepackage[capitalise]{cleveref}

\allowdisplaybreaks

\newcommand{\Eins}{\mathds{1}}%

\newcommand{\dif}{\mathrm{d}}%
\newcommand{\Nabla}{\vec{\nabla}}%
\newcommand{\fdif}{\operatorname{\delta}}%
\newcommand{\Fdif}[2]{\frac{\fdif\!#1}{\fdif\!#2}}%

\newcommand{\uu}{\vec{u}}

\newcommand{\rt}{(\vec{r},t)}

\newcommand{\ZT}[1]{\textquotedblleft#1\textquotedblright}%

\newcolumntype{Y}{>{\centering\arraybackslash}X}%     
\newcolumntype{Z}{>{\raggedright\arraybackslash}X}%   

\newlength{\myl}%
\newcommand{\SUM}[2]{{\setlength{\myl}{\widthof{$\displaystyle\sum_{#1}^{#2}$}*\real{0.5}-\widthof{$\displaystyle\sum$}*\real{0.5}}\sum_{#1}^{#2}\;\hspace{-\the\myl}}}% Summen in abgesetzten Gleichungen
\newcommand{\INT}[3]{\settowidth{\myl}{$\displaystyle\int_{#1}^{#2}$}{\int_{#1}^{#2}\;\;\;\hspace{-\the\myl}\dif #3}\,}% Integrale in abgesetzten Gleichungen
\newcommand{\TINT}[3]{\settowidth{\myl}{$\int_{#1}^{#2}$}{\int_{#1}^{#2}\!\ifthenelse{\equal{#1#2}{}}{}{\;\;\;\;\hspace{-\the\myl}}\dif #3}\,}%
\newcommand{\EINT}[3]{\settowidth{\myl}{$\int_{#1}^{#2}$}{\int_{#1}^{#2}\;\;\;\,\hspace{-\the\myl}\dif #3}\,}% Integrale in Exponenten
\newcommand{\CINT}[3]{\settowidth{\myl}{$\displaystyle\int_{#1}^{#2}$}{\oint_{#1}^{#2}\;\;\;\hspace{-\the\myl}\dif #3}\,}% geschlossene Integrale in abgesetzten Gleichungen

\begin{document}
\title{Passive and active field theories for disease spreading}

\author{Michael te Vrugt}
\affiliation{Institut f\"ur Theoretische Physik, Center for Soft Nanoscience, Westf\"alische Wilhelms-Universit\"at M\"unster, 48149 M\"unster, Germany}

\author{Julian Jeggle}
\affiliation{Institut f\"ur Theoretische Physik, Center for Soft Nanoscience, Westf\"alische Wilhelms-Universit\"at M\"unster, 48149 M\"unster, Germany}

\author{Raphael Wittkowski}
\email[Corresponding author: ]{raphael.wittkowski@uni-muenster.de}
\affiliation{Institut f\"ur Theoretische Physik, Center for Soft Nanoscience, Westf\"alische Wilhelms-Universit\"at M\"unster, 48149 M\"unster, Germany}

\begin{abstract}
The worldwide COVID-19 pandemic has led to a significant growth of interest in the development of mathematical models that allow to describe effects such as social distancing measures, the development of vaccines, and mutations. Several of these models are based on concepts from soft matter theory. Considerably less well investigated is the reverse direction, i.e., how results from epidemiological research can be of interest for the physics of colloids and polymers. In this work, we consider the SIR-DDFT model, a combination of the susceptible-infected-recovered (SIR) model from epidemiology with dynamical density functional theory (DDFT) from nonequilibrium soft matter physics, which allows for an explicit modeling of social distancing. We extend the SIR-DDFT model both from an epidemiological perspective by incorporating vaccines, asymptomaticity, reinfections, and mutations, and from a soft matter perspective by incorporating noise and self-propulsion and by deriving a phase field crystal (PFC) model that allows for a simplified description. On this basis, we investigate via computer simulations how epidemiological models are affected by the presence of non-reciprocal interactions. This is done in a numerical study of a zombie outbreak.
\end{abstract}
\maketitle

\section{Introduction}
The worldwide outbreak of the coronavirus disease 2019 (COVID-19), caused by the coronavirus SARS-CoV-2 \cite{WuEtAl2020,ZhouEtAl2020,WangHHG2020}, has inspired an enormous amount of research work on the spread of infectious diseases in the past years \cite{FrazierCDHJLSWZ2022,IHME2021,Estrada2020,Nesteruk2021}. A large portion of this research has focused on modeling the effects of various forms of interventions -- both nonpharmaceutical interventions such as contact restrictions \cite{DehningZSWNWP2020,MaierB2020,teVrugtBW2020,teVrugtBW2020b} and pharmaceutical interventions such as vaccination \cite{GiordanoCDFBBDNSCB2021,GrauerLL2020,HupertMDGAN2022} -- on the spread of the pandemic in order to develop optimal containment strategies. While the spread of COVID-19 is now mostly under control due to the successful development of vaccines \cite{MillerST2022}, research on modeling infectious diseases continues to be important for at least two reasons. First, the outbreak of further diseases (or mutations of older ones) is mostly a matter of time \cite{Quammen2013,Dodds2019}. Second, -- this aspect will be a focus of this article -- work on epidemic spreading has been fruitful also for other fields of research, such as soft matter physics \cite{teVrugtBW2020,teVrugtBW2020b,NorambuenaVG2020,ZhaoHR2022,GhoshCB2022,ForgacsLRHR2023}.

Epidemiological models range from compartmental models, such as the famous susceptible-infected-recovered (SIR) model \cite{KermackM1927}, which are very simple, to highly complex individual-based models \cite{WillemVBHB2017}, which allow to model an epidemic outbreak in a lot of detail. The \textit{SIR-DDFT} model \cite{teVrugtBW2020}, which is a combination of the SIR model with dynamical density functional theory (DDFT) \cite{teVrugtLW2020}, allows to combine the advantages of compartmental and individual-based models by allowing to model effects of social distancing explicitly within a simple (compared to individual-based models) coarse-grained field theory. Various extensions of the SIR-DDFT model have been developed or at least suggested, such as incorporating governmental interventions \cite{teVrugtBW2020b}, hydrodynamic interactions \cite{DuranK2021}, or determining model coefficients from Wi-Fi data \cite{YiXJ2021}. A numerical implementation is provided in Ref.\ \cite{JeggleW2021}, a brief review in Ref.\ \cite{Loewen2021b}.

The SIR-DDFT model is a paradigmatic example of a \textit{reaction-diffusion DDFT} (RDDFT) \cite{Lutsko2016,LutskoN2016}, which allows to simultaneously describe diffusion, chemical reactions, and particle interactions. In the past years, RDDFT has been widely used also outside of epidemiology \cite{teVrugtW2022}, in particular in active matter physics \cite{BleyDM2021,BleyHDM2021,MonchoD2020,BleyDM2021,AlstonPVB2022}.  Moreover, it has become an important tool in chemical engineering, where it has been applied to electrodes \cite{Liu2020}, metal corrosion \cite{ChenLL2022}, oxidation \cite{LiuL2020}, and reactions on catalytic substrates \cite{TangYZQXZ2021}. Related ideas are used in biophysical applications of DDFT \cite{WittmannNLSS2022,AlSaediHAW2018,ChauviereLC2012}. This suggests that a further development of the SIR-DDFT model should also be directed towards active matter in order to give it further significance also beyond the study of disease spreading, and that a deeper understanding of the SIR-DDFT model will have implications that go way beyond epidemiology.

In our first article on this topic \cite{teVrugtBW2020}, we have introduced the SIR-DDFT model and demonstrated its ability to model effects of social distancing. Our second article \cite{teVrugtBW2020b} has focused on epidemiological applications by considering effects of governmental interventions on the occurrence of multiple epidemic waves. The present third article complements the first two by considering not only questions of epidemiological interest, but also the importance of the SIR-DDFT model (and related theories) for physical research, in particular regarding active matter. We focus, in particular, on \textit{non-reciprocal interactions}, which have attracted enormous attention among soft matter physicists in the past years \cite{LoosKM2023,KreienkampK2022,SahaAG2020,LoosK2020,LinILHD2018,LoosK2020,LinILHD2018,MeredithMGCKvBZ2020,SchwarzendahlL2021,OuazanAG2023}. So far, they have not been incorporated into DDFT. Here, we develop a variant of the SIR-DDFT model (\textit{SZ-DDFT model}) that describes a zombie outbreak, a scenario which is governed by non-reciprocal interactions.

This article is structured as follows: In \cref{sirddftmodel}, we introduce the SIR-DDFT model. Section \ref{epextensions} presents extensions of the SIR-DDFT model that are motivated by epidemiological considerations. Extensions based on soft matter physics are presented in \cref{softextensions}. In \cref{simulations}, we show the results of numerical simulations. We conclude in \cref{conc}.

\section{\label{sirddftmodel}The SIR-DDFT model}
The starting point of our considerations is the very well known SIR model developed in Ref.\ \cite{KermackM1927} (based on earlier work \cite{Hamer1906,McKendrick1925}) and reviewed in Refs.\ \cite{Hethcote2000}, which describes the time evolution of the total number of susceptible ($\bar{S}$), infected ($\bar{I}$), and recovered ($\bar{R}$) persons. Susceptibles get infected at a rate $c_{\mathrm{eff}}\bar{I}$ with the effective contact rate $c_{\mathrm{eff}}$, and infected people recover at a rate $w$ and die at a rate $m$ (the SIR model with this extension is also referred to as \ZT{SIRD model} \cite{BergeLMMK2017}). These considerations lead to the dynamic equations
\begin{align}
\dot{\bar{S}} &= - c_{\mathrm{eff}}\bar{S}\bar{I}\label{s},\\  
\dot{\bar{I}} &= c_{\mathrm{eff}}\bar{S}\bar{I} - w\bar{I} - m \bar{I}\label{i},\\ 
\dot{\bar{R}} &= w\bar{I}\label{r}.
\end{align}
The SIR model given by these equations describes only the total number of persons in the respective compartments, but not their spatial distribution. This can be achieved by modeling not the total numbers, but the spatial densities $S$, $I$, and $R$ of susceptible, infected, and recovered persons. These are often assumed to obey the reaction-diffusion equations \cite{Noble1974}
\begin{align}
\partial_tS &= D_S \Nabla^2 S - cSI\label{dsr},\\  
\partial_tI &= D_I \Nabla^2 I +cSI - wI -mI\label{dir},\\ 
\partial_tR &= D_R \Nabla^2 R + wI\label{drr},
\end{align}
where $D_S$, $D_I$, and $D_R$ are the diffusion constants for susceptible, infected, and recovered persons and $c$ is the infection rate. This model is very useful for, e.g., describing the spread of animal diseases \cite{KeelingR2008}. If, however, we wish to describe the spread of diseases such as COVID-19 in a human society, the assumption implicitly underlying such reaction-diffusion equations, namely that all particles (persons) diffuse freely without affecting each other, is unrealistic due to the importance of social distancing, which can be thought of as a repulsive interaction. To tackle this problem, one can replace the diffusion terms in \cref{dsr,dir,drr} by DDFT terms, making use of the fact that DDFT is a generalization of the ordinary diffusion equation to interacting systems. This then gives an RDDFT model capable of describing epidemic spreading \cite{teVrugtBW2020}.

DDFT, developed in Refs.\ \cite{Evans1979,Munakata1989,Kawasaki1994,MarconiT1999} and reviewed in Ref.\ \cite{teVrugtLW2020}, is a dynamical theory for the one-body density $\rho$ of a fluid. Its central governing equation is given by
\begin{equation}
\partial_t \rho(\vec{r},t) = \Gamma\Nabla\cdot\bigg(\rho(\vec{r},t)\Nabla\Fdif{F}{\rho(\vec{r},t)}\bigg)
\label{ddft}
\end{equation}
with time $t$, position $\vec{r}$, mobility $\Gamma$, and free energy functional $F$. This functional is given by 
\begin{equation}
F= F_{\mathrm{id}} + F_{\mathrm{exc}}+ F_{\mathrm{ext}}
\label{freeenergy}
\end{equation}
with the ideal gas free energy
\begin{equation}
F_{\mathrm{id}} = k_\mathrm{B} T \INT{}{}{^dr}\rho(\vec{r})(\ln(\Lambda^3\rho(\vec{r}))-1)
\label{idealfreeenergy}
\end{equation}
with the number of spatial dimensions $d$, the Boltzmann constant $k_\mathrm{B}$, the temperature $T$, and the irrelevant thermal de Broglie wavelength $\Lambda$, the external contribution
\begin{equation}
F_{\mathrm{ext}}([\rho]) = \INT{}{}{^dr}\rho(\vec{r})U_1(\vec{r})
\end{equation}
with the external potential $U_1$, and the unknown excess free energy $F_{\mathrm{exc}}$ that incorporates particle interactions. 

In the present context, repulsive interactions represent the effects of social distancing and isolation. One has to distinguish here between general social distancing, which corresponds to all persons keeping a certain distance from each other, and self-isolation, which corresponds to the fact that persons who know that they are infected will put much more effort into staying away from other persons, generally by staying at home. On the level of the free energy, this is represented by including a  contribution for social distancing $F_{\mathrm{sd}}$ and a contribution for self-isolation $F_\mathrm{si}$. The excess free energy then reads
\begin{equation}
F_{\mathrm{exc}} = F_{\mathrm{sd}}+F_\mathrm{si}\label{free}
\end{equation}
with
\begin{align}
\begin{split}
F_{\mathrm{sd}} &= -\INT{}{}{^dr}\INT{}{}{^dr'} C_{\mathrm{sd}}e^{-\sigma_{\mathrm{sd}}(\vec{r}-\vec{r}')^2}\\
&\quad\:\! \bigg(\frac{1}{2}S(\vec{r},t)S(\vec{r}',t)+S(\vec{r},t)R(\vec{r}',t)+\frac{1}{2}R(\vec{r},t)R(\vec{r}',t)\bigg),\label{fsd}
\end{split}\\
\begin{split}
F_{\mathrm{si}} &= - \INT{}{}{^dr}\INT{}{}{^dr'} C_{\mathrm{si}}e^{-\sigma_{\mathrm{si}}(\vec{r}-\vec{r}')^2}I\rt\\
&\quad\:\! \bigg(\frac{1}{2}I(\vec{r}',t) 
+ S(\vec{r}',t)+R(\vec{r}',t)\bigg)\label{fsi},
\end{split}
\end{align}
where $C_{\mathrm{sd}}$ and $C_{\mathrm{si}}$ set the interaction strength and $\sigma_{\mathrm{sd}}$ and $\sigma_{\mathrm{si}}$ the interaction range. Thereby, we have assumed the interactions to take the form of a Gaussian soft-core repulsion, for which the excess free energy is well approximated by a mean-field approximation \cite{LouisBH2000}. 

Starting from \cref{dsr,dir,drr} and replacing the diffusion terms by the right-hand side of \cref{ddft} gives the \textit{SIR-DDFT model}
\begin{align}
\partial_tS &= \Gamma_S\Nabla\cdot\bigg( S\Nabla\Fdif{F}{S}\bigg) - cSI\label{sr},\\  
\partial_tI &= \Gamma_I\Nabla\cdot\bigg( I \Nabla\Fdif{F}{I}\bigg) +cSI - wI -mI\label{ir},\\ 
\partial_tR &=\Gamma_R\Nabla\cdot\bigg( R \Nabla\Fdif{F}{R}\bigg) + wI\label{rr}.
\end{align}

\section{\label{epextensions}Extensions motivated by epidemiology}
Since its initial development \cite{KermackM1927}, a huge number of extensions of the SIR model have been developed. Most of them can be incorporated pretty directly into the SIR-DDFT model. 
\subsection{Governmental interventions}
To make the list of epidemiological extensions complete, we start this section by showing an extension that was already derived in Ref.\ \cite{teVrugtBW2020b}. The aim here is to incorporate the fact that nonpharmaceutical interventions such as contact restrictions are not present in the same intensity at all times, but are imposed and lifted depending on the current infection numbers. In the SIR-DDFT model, this means that the interactions are time-dependent. For this purpose, we add (based on the rectangular hysteresis model by \citet{ChladnaKRR2020}) dynamic equations for the coefficients $C_{\mathrm{sd}}$ and $C_{\mathrm{si}}$ that are given by
\begin{equation}
\dot{C}_i(t) =
\begin{cases}
\alpha (C_{i,0} - C_i (t)) &\text{if }(\bar{I}(\tau) < \bar{I}_\mathrm{start} \forall \tau \in [0,t])\\ 
& \text{or } (\exists t_1 \in [0,t] \text{ such that }\\ 
& \bar{I}(t_1) \leq \bar{I}_\mathrm{stop}  \text{ and }\\ 
& \bar{I}(\tau) < \bar{I}_\mathrm{start} \forall \tau \in (t_1,t]),\\
\alpha (C_{i,1} - C_i (t)) &\text{if }\exists t_1 \in [0,t] \text{ such that}\\ 
& \bar{I}(t_1) \geq \bar{I}_\mathrm{start}\text{ and } \\ 
& \bar{I}(\tau) > \bar{I}_\mathrm{stop} \forall \tau \in (t_1,t]
\end{cases}
\label{cr}%
\end{equation}
with $i=\mathrm{sd}, \mathrm{si}$. Here, $C_{i,1}$ and $C_{i,0}$ are the values that the parameter $C_i$ approaches in the presence and absence of a shutdown, respectively, and $\alpha$ measures the rate at which these values are approached.  The physical idea is that the start and end of a shutdown is triggered if the number of infected persons $\bar{I}$ passes threshold values $\bar{I}_\mathrm{start}$ and $\bar{I}_\mathrm{stop}$, respectively.

\subsection{Vaccination}
The rapid development of safe and effective vaccines \cite{MillerST2022} has had a tremendous impact on containing the spread of the COVID-19 pandemic. During the early stages of the vaccination campaign, the supply of vaccines has been very limited. In such contexts, it is important to develop strategies for distributing them in order to use the available vaccines efficiently \cite{GrauerLL2020}. 

In the SIR-DDFT model, vaccination can be incorporated in two basic ways. The first and simpler one is to assume that vaccination has the same effect as recovery, such that vaccination simply transfers persons from the susceptible to the recovered compartment \cite{KopfovaNRR2020,EhrhardtGK2019}. The second one is to introduce a fourth compartment that contains vaccinated persons \cite{LaguzetT2015,ChauhanMD2014}. In a theory such as the SIR-DDFT model that is based on partial rather than ordinary differential equations, it is desirable to keep the number of compartments small in order to reduce the numerical cost. Therefore, we use the first variant here. Hence, we assume that vaccination essentially transfers a person directly from the S- to the R-compartment since vaccinated persons (just like recovered ones) are immune. Let $v$ be the vaccination rate, which can depend on space and time (e.g., because vaccines are first introduced in a certain region or because vaccine skepticism is more prevalent in some regions than in others). In this case, the SIR-DDFT model with vaccination reads
\begin{align}
\partial_tS &= \Gamma_S\Nabla\cdot\bigg( S\Nabla\Fdif{F}{S}\bigg) - cSI - vS,\\  
\partial_tI &= \Gamma_I\Nabla\cdot\bigg( I \Nabla\Fdif{F}{I}\bigg) +cSI - wI -mI,\\ 
\partial_tR &=\Gamma_R\Nabla\cdot\bigg( R \Nabla\Fdif{F}{R}\bigg) + wI +vS.
\end{align}
If (as it is the case for COVID-19) the vaccine does not lead to full immunity, this model can be combined with an extension allowing for reinfections (see \cref{reinf}).

\subsection{\label{expo}Exposed and asymptomatic persons}
The standard SIR-DDFT model assumes that susceptibles immediately become infected upon contact with an infected person, and that all infected persons then exhibit the same repulsive interaction. This assumption is unphysical for two important reasons: 
\begin{enumerate}
    \item The disease has a certain incubation period, i.e., one does not immediately become infectious after having caught the disease. 
    \item While some people become severely ill, others have mild or no symptoms (but can still infect others). 
\end{enumerate}
The existence of asymptomatic individuals that can infect others and the fact that people can be infectious already before developing symptoms has been of enormous importance for the spread of COVID-19. Consequently, introducing exposed and asymptomatic compartments is a relatively natural extension of the SIR model \cite{AdekolaAEOA2020}. In the context of interaction modeling in the SIR-DDFT model, this distinction is important because only symptomatic individuals will self-isolate. Asymptomatic infected persons, on the other hand, will simply interact with others in exactly the same way as susceptible ones. Physically speaking, the interaction potential cannot distinguish between susceptible and asymptomatic infected persons. 

Following \citet{GrauerLL2020}, we therefore add two fields to the SIR-DDFT model. First, $E$ is the density of exposed persons. Second, $A$ is the density of asymptomatic infected persons. Upon contact with an infected person, a susceptible person first becomes exposed at a rate $c$. It then becomes asymptomatic (following the observation that, for COVID-19, persons are infectious before they are symptomatic) at a rate $k_1$. Then, it either stays asymptomatic until it recovers or it becomes symptomatic. This is incorporated by transitions from the asymptomatic to the infectious compartment, taking place at a rate $k_2$, and transitions from the asymptomatic to the recovered compartment, taking place at a rate $w_1$. Together, these assumptions give the dynamic equations
\begin{align}
\partial_tS &= \Gamma_S\Nabla\cdot\bigg( S\Nabla\Fdif{F}{S}\bigg) - c S(A+I),\\
\partial_t E &= \Gamma_I\Nabla\cdot\bigg( I \Nabla\Fdif{F}{I}\bigg) +cS(A+I) - k_1  E,\\ 
\partial_t A &= \Gamma_I\Nabla\cdot\bigg( I \Nabla\Fdif{F}{I}\bigg) + k_1 E - (w_1+k_2) A,\\ 
\partial_tI &= \Gamma_I\Nabla\cdot\bigg( I \Nabla\Fdif{F}{I}\bigg)+ k_2 A - (w+m) I,\\ 
\partial_tR &=\Gamma_R\Nabla\cdot\bigg( R \Nabla\Fdif{F}{R}\bigg) + w_1 A + w I.
\end{align}

\subsection{\label{reinf}Reinfections}
The SIR-DDFT model was developed in 2020, when little was known about the likelihood of a re-infection after a person has recovered from COVID-19. It assumes, as done in the simple SIR model, that a person that has recovered from the disease is completely immune. Today, it is known that reinfections with COVID-19 are possible \cite{WestEN2021,PradoBGMGRGTBC2021}. In SIR-type models, reinfections can be incorporated in different ways depending on the properties of the disease. For example, the immunity acquired after recovery may be only partial, or it may decrease over time \cite{GomesWM2004}.

What an extension of the SIR-DDFT model incorporating reinfections has to look like thus depends on the disease in question. We consider here the case of COVID-19, for which the current knowledge about reinfection probability is reviewed in Refs.\ \cite{SteinEtAl2023,OMurchuBCDKOHR2022}. The probability of reinfection is low for the wild type and for Alpha, Beta, and Delta variants. For the Omicron variant, it is higher, but a previous infection still provides a reasonable protection against severe disease \cite{SteinEtAl2023}. Here, we simply assume that there is a constant reinfection rate $c_R$ by which recovered persons can get infected when encountering infected individuals. This gives
\begin{align}
\partial_tS &= \Gamma_S\Nabla\cdot\bigg( S\Nabla\Fdif{F}{S}\bigg) - cSI,\\  
\partial_tI &= \Gamma_I\Nabla\cdot\bigg( I \Nabla\Fdif{F}{I}\bigg) +cSI + c_R R I - wI -mI,\\ 
\partial_tR &=\Gamma_R\Nabla\cdot\bigg( R \Nabla\Fdif{F}{R}\bigg) + wI -c_R R I. 
\end{align}
We do not incorporate here the fact that reinfections appear to be milder \cite{QureshiBHLHS2022}, which would imply that $m$ is different for the first and the second infection. The reason is that doing so would make the model significantly more complicated while providing little additional epidemiologically relevant information.

\subsection{Mutations}
A particular central factor for the occurrence of re-infections are mutations, in particular so-called \ZT{immune escape variants} such as the Omicron variant \cite{HuPCWCWTH2022}. The emergence of new variants has had a major impact on the course of the COVID-19 pandemic, which has motivated also the development of models specifically aiming to incorporate mutations \cite{SchwarzendahlGLL2021}.

In the SIR-DDFT model, we can incorporate a mutation by adding a field $I_\mathrm{M}$ that describes the density of persons infected with a mutation, as well as a field $R_\mathrm{M}$ for persons that have recovered from an infection with the mutation. (A similar approach was used in Ref.\ \cite{SchwarzendahlGLL2021} for a normal SIR model.) We moreover make the following assumptions:
\begin{itemize}
\item Susceptibles are infected with a rate $c$ when encountering someone infected with the wild type and with a rate $c_\mathrm{M}$ when encountering someone infected with a mutation. 
\item Recovered persons are immune against the variant they have recovered from. Someone who has been infected with the mutation is additionally immune against the wild type, but not vice versa. Therefore, recovered persons can be infected by the mutation at a rate $c_{\mathrm{MR}}$. One cannot be infected with both variants at the same time.
\item Persons infected with the wild type recover at a rate $w$ and die at a rate $m$. Persons infected with the mutation recover at a rate $w_\mathrm{M}$ and die at a rate $m_\mathrm{M}$.
\end{itemize}
With these assumptions, the SIR-DDFT model with mutations reads
\begin{align}
\partial_tS &= \Gamma_S\Nabla\cdot\bigg( S\Nabla\Fdif{F}{S}\bigg) - cSI - c_\mathrm{M} S I_\mathrm{M},\\  
\partial_tI &= \Gamma_I\Nabla\cdot\bigg( I \Nabla\Fdif{F}{I}\bigg) +cSI - wI -mI,\\ 
\partial_tI_\mathrm{M} &= \Gamma_{I_\mathrm{M}}\Nabla\cdot\bigg( I_\mathrm{M} \Nabla\frac{\delta F}{\delta I_\mathrm{M}}\bigg) +c_\mathrm{M} S I_\mathrm{M} + c_{\mathrm{MR}} R I_\mathrm{M} \notag \\&\quad - w_\mathrm{M} I_\mathrm{M} -m_\mathrm{M} I_\mathrm{M},\\ 
\partial_tR &=\Gamma_R\Nabla\cdot\bigg( R \Nabla\Fdif{F}{R}\bigg) + wI - c_{\mathrm{MR}} R I_\mathrm{M},\\
\partial_tR_\mathrm{M} &=\Gamma_{R_\mathrm{M}}\Nabla\cdot\bigg( R_\mathrm{M} \Nabla\frac{\delta F}{\delta R_\mathrm{M}}\bigg) + wI +w_\mathrm{M} I_\mathrm{M}.
\end{align}
Here, $\Gamma_{I_\mathrm{M}}$ and $\Gamma_{R_\mathrm{M}}$ are the mobilities of persons that are infected with the mutation or have recovered from it, respectively.

\section{\label{softextensions}Extensions motivated by soft matter physics}
\subsection{Noise}
The SIR-DDFT model is a deterministic theory that originates from combining the deterministic SIR model with deterministic DDFT. Interestingly, both theories also exist in stochastic variants (see Refs.\ \cite{HespanhaCCEY2021,GreenwoodG2009} for reviews of stochastic SIR models and Ref.\ \cite{teVrugtLW2020} for a review of stochastic DDFT), such that it is natural to ask how stochasticity can be incorporated into the SIR-DDFT model. In fact, the incorporation of noise has already been suggested in Ref.\ \cite{DuranK2021}, although it has not been done explicitly.

The early days of DDFT saw a coexistence between deterministic \cite{MarconiT1999,MarconiT2000} and stochastic \cite{Dean1996,Kawasaki1994} variants, and it was a matter of debate which form is the \ZT{correct} one. Nowadays, following work by \citet{ArcherR2004}, it is generally understood that deterministic DDFT describes the ensemble-averaged density (which in this context is the average of the density over all realizations of the noise in the Langevin equations that describe the motion of individual particles), whereas stochastic DDFT describes either the microscopic density operator or a spatially coarse-grained density. From this point of view, it makes sense that the SIR-DDFT model, which was introduced as a theory for the ensemble-averaged density \cite{teVrugtBW2020}, contains no noise terms as these are averaged over. 

Nevertheless, in a real experiment (or pandemic), one observes not an ensemble average of the density of all possible time evolutions, but rather a spatial average of the density \cite{ArcherR2004}. Thus, from a practical point of view, it is desirable to take noise terms into account. This can be done in a straightforward way by simply replacing the deterministic DDFT terms in \cref{sr,ir,rr} by the ones known from stochastic DDFT \cite{Kawasaki1994}. The result is
\begin{align}
\partial_tS &= \Gamma_S\Nabla\cdot\bigg( S\Nabla\Fdif{F}{S}\bigg) + \Nabla\cdot(\sqrt{2\Gamma_S k_\mathrm{B} T_S S}\vec{\eta}_S) - cSI,\label{noise1}\\  
\partial_tI &= \Gamma_I\Nabla\cdot\bigg( I \Nabla\Fdif{F}{I}\bigg) + \Nabla\cdot(\sqrt{2\Gamma_I k_\mathrm{B} T_I I}\vec{\eta}_I)\notag\\&\quad +cSI - wI -mI\label{noise2},\\ 
\partial_tR &=\Gamma_R\Nabla\cdot\bigg( R \Nabla\Fdif{F}{R}\bigg) + \Nabla\cdot(\sqrt{2\Gamma_R k_\mathrm{B} T_R R}\vec{\eta}_R)+ wI\label{noise3}
\end{align}
with the temperatures $T_i$, where the noise $\vec{\eta}_i$ has the properties
\begin{align}
\braket{\vec{\eta}_i\rt}&=\vec{0},\\
\braket{\vec{\eta}_i\rt\otimes\vec{\eta}_j(\vec{r}',t')}&=\Eins\delta_{ij}\delta(\vec{r}-\vec{r}')\delta(t-t').
\end{align}
Here, $\braket{\cdot}$ denotes an ensemble average, $\otimes$ is a dyadic product, $\Eins$ is the unit matrix, and $\delta$ is the Dirac delta distribution.

The stochastic SIR-DDFT model given by \cref{noise1,noise2,noise3} should be understood as a description of the actual (possibly spatially averaged) densities, whereas the deterministic model usually considered gives the ensemble-averaged densities. Note that integrating \cref{noise1,noise2,noise3} over space still gives the deterministic SIR model. Mathematically, this is due to the fact that the noise terms are still written as the divergence of a conserved current (the densities are conserved in the absence of infection dynamics). Physically, this is due to the fact that we are considering here the stochasticity of the spatial motion, which is invisible on the level of the spatially averaged SIR model. Consequently, \cref{noise1,noise2,noise3} can be thought of as a combination of stochastic DDFT with the deterministic SIR model. Nevertheless, since the spatial distribution of the persons determines the effective infection rate $c_{\mathrm{eff}}$ \cite{teVrugtBW2020}, fluctuations of the density fields will in practice also induce fluctuations on the SIR level. Combinations of the stochastic SIR model with deterministic or stochastic DDFT represent further possible extensions.

\subsection{Active matter}
Active matter is characterized by a continuous inflow of energy at a local level, typically with the consequence that the particles exhibit directed motion \cite{MarchettiJRLPRS2013,BechingerdLLRVV2016}. While active particles can also be realized artificially, for example by ultrasound-generated propulsion \cite{VossW2018,VossW2020}, the most generic example for active particles are biological organisms such as swimming bacteria or flying birds. Biological organisms, of course, are also where infectious diseases are spreading, and the fact that systems of biological organisms constitute active matter might have important influences on the dynamics of a pandemic. Consequently, it is not surprising that a lot of research has been devoted to studying the connection between disease spreading and active matter (see, for example, Refs.\ \cite{NorambuenaVG2020,ZhaoHR2022,ForgacsLRHR2022,GhoshCB2022,ForgacsLRHR2023,LibalFNRHR2022}). 

The SIR-DDFT model -- just like the reaction-diffusion SIR model it is derived from -- assumes the motion of humans to be essentially described by the motion of passive Brownian particles. This can be true at most approximately since human motion arises not from being kicked around by a thermal fluid, but from the conversion of internal energy into directed motion. In other words, humans are active particles. The dynamics of active particles can differ in many interesting ways from that of passive ones, and this difference may be of importance for the spread of diseases. Likewise, the dynamics of chemical reactions in a system with steric interactions (described by RDDFT) might be affected by the fact that some reactants are transported by active processes. Consequently, an extension of the SIR-DDFT model to the active case is desirable. 

Theoretical models of (overdamped) active particles typically take them to be described not just by their position $\vec{r}$, but also by their orientation vector $\uu$ that in two spatial dimensions can be parametrized by an angle $\phi$. The orientation vector describes the direction of self-propulsion. For an active DDFT, this implies that the density depends not only on $\vec{r}$, but also on $\phi$. The governing equation gets additional terms describing rotational diffusion (change of the particle orientation) and self-propulsion, respectively. An active DDFT was first derived by \citet{WensinkL2008}. Later work extended this theory to particles with arbitrary shape \cite{WittkowskiL2011} or microswimmers \cite{MenzelSHL2016}. 

Assuming isotropic translational diffusion, the active DDFT equation reads \cite{WensinkL2008,MenzelOL2014}
\begin{equation}
\begin{split}
\partial_t \rho &= \Gamma \Nabla\cdot\bigg(\rho\Nabla\Fdif{F}{\rho}\bigg)+D_\mathrm{R} \beta\partial_\phi \bigg(\rho\partial_\phi\Fdif{F}{\rho}\bigg)\\
&\quad -v_0\Nabla\cdot(\rho\uu),
\end{split}
\label{activeddft}
\end{equation}
where $D_\mathrm{R}$ is the rotational diffusion coefficient, $\beta$ the rescaled inverse temperature, $v_0$ the self-propulsion velocity, and $\uu(\phi)=(\cos(\phi),\sin(\phi))^\mathrm{T}$ the particle orientation. If we assume the particles to be active -- in our context, if we assume that humans are not passively diffusing, but walking around with a velocity $v$ -- then we have to replace the diffusion term in the reaction-diffusion SIR model not with the passive DDFT \eqref{ddft}, but with the active DDFT \eqref{activeddft}.

In the active case, we also have to be a lot more careful with the reaction terms. Let us consider, as an example, the term $cSI$ in the governing equation for $I$. In a passive system, we have
\begin{equation}
\partial_tI(\vec{r},t) = c S(\vec{r},t)I(\vec{r},t)+\dotsb
\end{equation}
(we write the arguments of the fields for the moment for illustration). Physically, this means that rate of new infections at position $\vec{r}$ is proportional to the number of susceptible and infected persons at this position, since susceptibles become infected when meeting an infected person. A naive generalization to the active case would be
\begin{equation}
\partial_tI(\vec{r},\phi,t) = c S(\vec{r},\phi,t)I(\vec{r},\phi,t)+\dotsb,
\end{equation}
which assumes that the structure of the reaction-diffusion model is unaffected by the presence of additional orientational degrees of freedom. This, however, is unrealistic. For the question whether a susceptible person can be infected if they are at the same position as an infected person, the orientation of the infected person should not matter (at least to a first approximation). To get an additional infected person with orientation $\phi$ at position $\vec{r}$, we require a susceptible person with orientation $\phi$ at position $\vec{r}$ and an infected person \textit{with any orientation} at position $\vec{r}$. This gives
\begin{equation}
\partial_tI(\vec{r},\phi,t) = c S(\vec{r},\phi,t)\INT{0}{2\pi}{\phi'}I(\vec{r},\phi',t)+\dotsb.
\end{equation}
A more sophisticated model could take into account that an infection is more likely if the persons are looking at each other. More generally, the infection probability $c$ might depend on $\phi-\phi'$. This leads to
\begin{equation}
\partial_tI(\vec{r},\phi,t) =\INT{0}{2\pi}{\phi'}c(\phi-\phi')S(\vec{r},\phi,t)I(\vec{r},\phi',t)+\dotsb,
\end{equation}
which can (now dropping arguments again) be written as
\begin{equation}
\partial_t I = (c \star_\phi I) S
\end{equation}
with the orientational convolution $\star_\phi$. Note that these considerations are relevant not only in an epidemiological context, but also if we wish to model actual chemical reactions in which active particles (or, more generally, particles with orientational degrees of freedom) are involved. When deriving a reaction-diffusion model for such systems, it has to be taken into account whether and how the particles' orientation affects the reactions they undergo. This further emphasizes the relevance of the present study for soft matter physics.

We thus arrive at the \textit{active SIR-DDFT model}, given by
\begin{align}
\begin{split}
\partial_tS &= \Gamma_S\Nabla\cdot\bigg( S\Nabla\Fdif{F}{S}\bigg) + D_{\mathrm{R},S}\beta_S\partial_\phi\bigg(S\partial_\phi\Fdif{F}{S}\bigg)\\
&\quad - v_S\Nabla\cdot(S\uu) - (c \star_\phi I) S,
\end{split}\\
\begin{split}
\partial_tI &= \Gamma_I\Nabla\cdot\bigg(I\Nabla\Fdif{F}{I}\bigg) + D_{\mathrm{R},I}\beta_I\partial_\phi\bigg(I\partial_\phi\Fdif{F}{I}\bigg)\\
&\quad - v_I\Nabla\cdot(I\uu) - (c \star_\phi I) S - wI - mI,
\end{split}\\
\begin{split}
\partial_tR &= \Gamma_R\Nabla\cdot\bigg(R\Nabla\Fdif{F}{R}\bigg) + D_{\mathrm{R},R}\beta_R\partial_\phi\bigg(R\partial_\phi\Fdif{F}{R}\bigg)\\
&\quad - v_R\Nabla\cdot(R\uu) +wI.
\end{split}
\end{align}
Note that we have allowed the rotational diffusion coefficients $D_{\mathrm{R},i}$, the rescaled inverse temperatures $\beta_i$, and the self-propulsion velocities $v_i$ to be different for the different fields. This can, for example, be a consequence of ill persons walking slower than noninfected ones.

\subsection{Phase field crystal model}
The governing equations of DDFT can be quite difficult to solve in practice, in particular due to the convolution in the interaction term. Therefore, it is desirable to have available simpler models that still capture the same essential physics. This requirement is satisfied by \textit{phase field crystal} (PFC) models. After their phenomenological introduction \cite{ElderKHG2002,ElderG2004,BerryGE2006}, it has been found that they can be derived as an approximation to DDFT \cite{ElderPBSG2007,vanTeeffelenBVL2009}. This derivation is discussed in detail in Refs.\ \cite{ArcherRRS2019,teVrugtLW2020,teVrugtHKWT2022}. PFC models also allow to model mixtures \cite{HuangEP2010,TahaDMEH2019,ElderKHG2002,HollAT2020,RobbinsATK2012,AlaimoV2018,HollAGKOT2020,teVrugtHKWT2022} and active matter \cite{MenzelL2013,MenzelOL2014,teVrugtJW2021,OphausGT2018,OphausKGT2020,OphausKGT2020b,teVrugtHKWT2022}. They are reviewed in Ref.\ \cite{EmmerichEtAl2012}.

In most cases, the order parameter of PFC models is given by the dimensionless deviation of the density from its mean value. In the present case, however, this would not be a convenient choice since this would make the reaction terms unnecessarily complicated. Therefore, we simply use $S$, $I$, and $R$ as order parameter fields also for the PFC model. Starting from the SIR-DDFT model, we make three standard approximations:
\begin{enumerate}
\item We replace the expression $\Nabla\cdot\phi\Nabla$ (with $\phi=S,I,R$) in front of $\delta F/\delta \phi$ by $\hat{\rho}\Nabla^2$ with a reference density $\hat{\rho}$. This approximation is straightforward in standard DDFT where one can simply choose the average density, but is a little more tricky in RDDFT since the individual densities are not conserved and the density of, e.g., susceptibles can deviate quite a lot from any reference value one might choose and moreover depends on time. For the reference density, we therefore here use the mean population density $\hat{\rho}=N_0/A$ with the initial total number of persons $N_0$ and the domain area $A$. Thereby, we ensure that $\hat{\rho}$ is constant.
\item We make a Taylor expansion for the ideal gas free energy \eqref{idealfreeenergy} around $\hat{\rho}$ up to fourth order in $S$, $I$, and $R$.
\item The nonlocal excess free energy given by \cref{free,fsd,fsi} is made local by a gradient expansion up to fourth order. 
\end{enumerate}

As a result, we obtain the \textit{SIR-PFC model}, which is given by 
\begin{align}
\partial_tS &= \Gamma_S\hat{\rho}\Nabla^2\Fdif{F}{S} - cSI,\label{srp}\\  
\partial_tI &= \Gamma_I\hat{\rho}\Nabla^2\Fdif{F}{I} +cSI - wI -mI,\label{irp}\\ 
\partial_tR &=\Gamma_R\hat{\rho}\Nabla^2\Fdif{F}{R} + wI\label{rrp}.
\end{align}
The Taylor-expanded ideal gas free energy reads (ignoring irrelevant zeroth- and first-order terms)
\begin{equation}
F_{\mathrm{id}}=\sum_{\phi=S,I,R}\beta_\phi^{-1}\hat{\rho}\INT{}{}{^dr}\frac{3\phi^2}{2\hat{\rho}^2}-\frac{\phi^3}{2\hat{\rho}^3}+\frac{\phi^4}{12\hat{\rho}^4}.
\label{idealfreeenergypfc}
\end{equation}
Equation \eqref{idealfreeenergypfc} has a different form than the ideal gas free energy in a standard PFC model \cite{EmmerichEtAl2012}. This is simply a consequence of the fact that we work with the density rather than the density deviation.

Equations \eqref{fsd} and \eqref{fsi} simplify to
\begin{align}
\begin{split}
F_{\mathrm{sd}} &= -\INT{}{}{^dr} \frac{1}{2}(C_{\mathrm{sd}}^{(0)}S^2 + C_{\mathrm{sd}}^{(2)}S\Nabla^2S + C_{\mathrm{sd}}^{(4)}S\Nabla^4S\\
&\quad\:\! +2C_{\mathrm{sd}}^{(0)}SR +2 C_{\mathrm{sd}}^{(2)}S\Nabla^2R + 2C_{\mathrm{sd}}^{(4)}S\Nabla^4R\\
&\quad\:\!+C_{\mathrm{sd}}^{(0)}R^2 + C_{\mathrm{sd}}^{(2)}R\Nabla^2R+C_{\mathrm{sd}}^{(4)}R\Nabla^4 R),\label{fsdpfc}
\end{split}\\
\begin{split}
F_{\mathrm{si}} &= - \INT{}{}{^dr}\frac{1}{2}(C_{\mathrm{si}}^{(0)}I^2 + C_{\mathrm{si}}^{(2)}I\Nabla^2I + C_{\mathrm{si}}^{(4)}I\Nabla^4I\\
&\quad\:\! +2C_{\mathrm{si}}^{(0)}IS +2 C_{\mathrm{si}}^{(2)}I\Nabla^2S + 2C_{\mathrm{si}}^{(4)}I\Nabla^4S\\
&\quad\:\! +2C_{\mathrm{si}}^{(0)}IR +2 C_{\mathrm{si}}^{(2)}I\Nabla^2R + 2C_{\mathrm{si}}^{(4)}I\Nabla^4R)\label{fsipfc}
\end{split}
\end{align}
with the parameters
\begin{align}
C_{i}^{(0)} =&\frac{\pi}{\sigma_{i}}C_{i},\label{c0}\\
C_{i}^{(2)} =&\frac{\pi}{4\sigma_{{i}^2}}C_{i},\label{c2}\\
C_{i}^{(4)} =&\frac{\pi}{32\sigma_{i}^3}C_{i}\label{c4}
\end{align}
and $i = \mathrm{sd}, \mathrm{si}$. Note that \cref{fsdpfc,fsipfc} hold in any spatial dimension, whereas \cref{c0,c2,c4} hold only in $d=2$ dimensions. Moreover, the free energy of the SIR-PFC model does not have the familiar Swift-Hohenberg-type \cite{SwiftH1977} form. This is a direct consequence of the fact that we do not shift or rescale the density fields. Doing so would make the nonconserved part of the dynamics, which is not present in standard PFC models, significantly more complicated. 

This also indicates that Swift-Hohenberg free energies are generally less appropriate if a PFC model is used to study systems with chemical reactions, an observation that is of interest also for applications in chemical engineering or biochemistry. Chemical reaction networks are important, for instance, also in the development of intelligent materials \cite{KasparRvdWWP2021} or in biological systems \cite{OuazanAG2023}, and knowing how to extend PFC models to interacting systems of reacting species is therefore useful also for these fields of research.

\subsection{Zombie outbreak}
We finally consider an epidemic scenario that is somewhat different from that of virus spreading, namely a zombie outbreak. Zombies, originating from Haitian folk belief, have developed into a very common motive in popular culture such as novels, movies, or video games. Moreover, zombie outbreaks have been the subject of various mathematical modeling studies since they provide an interesting case study for epidemiological models \cite{VerranJA2014,WatsonHFE2014,VerranR2018,MunzHIS2009,LibalFNRHR2022}. See Ref.\ \cite{SmithS2011} for an overview over zombie-related research.

While zombies have not been a major public health concern in the past years, they are extremely interesting in the context of this work from a physical point of view. While it can be assumed that, in a COVID-19 outbreak, infected persons try to keep a distance from noninfected ones just as they are keeping one from them (reciprocal interactions), zombies exhibit a different behavior: they actively attack non-infected humans and try to bite them, whereas humans will run away from zombies in order to avoid being killed. Thus, the interaction between humans and zombies is \textit{non-reciprocal}. Non-reciprocal couplings have attracted a lot of interest in recent years \cite{LoosK2020,LinILHD2018}, including in the context of field theories \cite{LoosK2020,LinILHD2018} and predator-prey models \cite{MeredithMGCKvBZ2020,SchwarzendahlL2021}. Thus, studying a zombie outbreak in an SIR-DDFT model represents an interesting contribution to modern active matter physics.

As a starting point, we use the susceptible-zombie-removed (SZR) model \cite{AlemiBMS2015}, which is given by
\begin{align}
	\dot{\bar{S}} &= - b_{\mathrm{eff}}\bar{S}\bar{Z},\label{szrmodel1}\\  
	\dot{\bar{Z}} &= (b_{\mathrm{eff}}-\kappa_{\mathrm{eff}})\bar{S}\bar{Z},\label{szrmodel2}\\ 
	\dot{\bar{R}} &= \kappa_{\mathrm{eff}}\bar{S}\bar{Z}.\label{szrmodel3}
\end{align}
Here, $b_{\mathrm{eff}}$ stands for the effective bite parameter (rate at which zombies bite humans), $\kappa_{\mathrm{eff}}$ is the effective kill parameter (rate at which humans kill zombies), $\bar{Z}$ is the total number of zombies, and $\bar{R}$ is the number of removed individuals, i.e., the number of killed zombies. Note that, although the $\bar{R}$ compartment plays a similar mathematical role as in the SIR model, the physical interpretation here is different since the only way to be removed from the zombie population is through death \cite{WitkowskiB2013}.

Similar as in the SIR-DDFT model, we now consider the spatial densities $S$ and $Z$. We can ignore the removed compartment here since the corresponding field would simply describe the spatial distribution of zombie corpses. (While accumulations of zombie corpses are certainly unpleasant, we can safely assume them to be irrelevant for the overall dynamics.) By basing our considerations on the SZR model, we make two central approximations. First, we neglect the effects of suicides that susceptibles commit in order to avoid becoming a zombie. This is, as discussed in Ref.\ \cite{WitkowskiB2013}, a good approximation for the movie \textit{Shaun of the Dead} \cite{WrightP2004}, where no suicides happen, but more problematic for other zombie movies. Second, we ignore the effects of exposure (see \cref{expo}) and simply assume that a bitten person immediately becomes a zombie. This can be justified by the short half life of exposed persons (about 30 minutes for \textit{Shaun of the Dead} \cite{AlemiBMS2015}). We make this second approximation because it is not fully clear how to accommodate their behavior within our model -- depending on their character traits and on whether exposed persons are aware of their fate, they may be running away from susceptibles or from zombies, will continue to kill zombies or not, or will even be killed by susceptibles. 

Denoting the local bite and kill parameters by $b$ and $\kappa$, respectively, we propose the \textit{SZ-DDFT model}, which is given by 
\begin{align}
	\begin{split}
		\partial_tS &= D_S\Nabla^2 S - \Gamma_S\Nabla\cdot\big(S \Nabla ( C_{\mathrm{sz}}K_{\mathrm{sz}}\star Z)\big) -  bS Z\label{zombie1},
	\end{split}\raisetag{1.5em}\\  
	\begin{split}
		\partial_t Z &= D_Z\Nabla^2 Z + \Gamma_Z\Nabla\cdot\big( Z\Nabla(C_{\mathrm{zs}}K_{\mathrm{zs}}S\big)\big)\\&\quad +(b-\kappa) S Z \label{zombie2},
	\end{split}\raisetag{1.5em}
\end{align}
Here, $D_S = \beta_S \Gamma_S$ is the diffusion constant for the susceptibles, $D_Z = \beta_Z \Gamma_Z$ is the diffusion constant for the zombies (depending on their inverse rescaled temperature $\beta_Z$ and their mobility $\Gamma_Z$), $C_{\mathrm{sz}}$ is the strength of the \textit{repulsive} force that the zombies exert on the susceptibles, $C_{\mathrm{zs}}$ is the strength of the \textit{attractive} force the humans exert on the zombies, $\star$ denotes a spatial convolution, and 
\begin{align}
	K_i(\vec{r})=e^{-\sigma_i\vec{r}^2}
\end{align}
with the interaction ranges $\sigma_i$ and $i=\mathrm{sz},\mathrm{zs}$ are the kernels. The constants $C_{\mathrm{sz}}$ and $C_{\mathrm{zs}}$ have to be negative for physical reasons, but can be different (depending on how hungry the zombies and how scared the susceptibles are). Essentially, the SZ-DDFT model is obtained from the SIR-DDFT model by dropping the field $R$, changing the reaction terms, and flipping the sign of the interaction term in the dynamic equation for $Z$ while keeping the sign in the dynamic equation for $I$. Thereby, the model becomes non-reciprocal. Consequently, the zombie model constitutes another active field theory for epidemic spreading.

\begin{figure*}
    \centering\includegraphics[width=\linewidth]{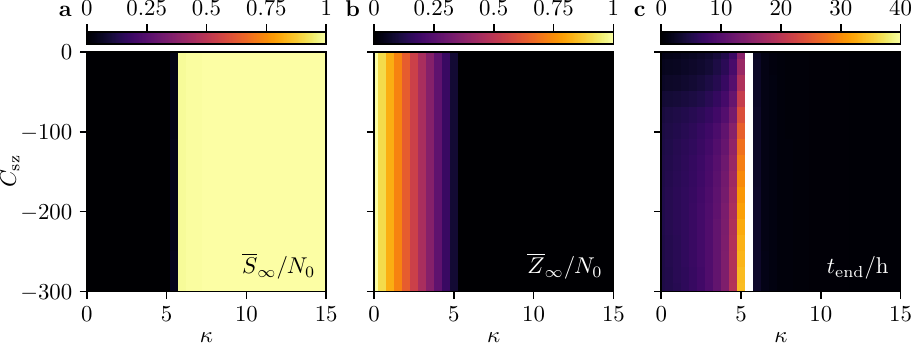}
    \caption{Phase diagram of the SZ-DDFT model, showing \textbf{(a)} the normalized final number of susceptibles $\bar{S}_\infty/N_0$, \textbf{(b)}, the normalized final number of zombies $\bar{Z}_\infty/N_0$, and \textbf{(c)} the time $t_\mathrm{end}$ it takes for the battle between susceptibles and zombies to be over as a function of $\kappa$ (kill rate) and $C_{\mathrm{zs}}$ (strength of the repulsive force acting on susceptibles).}
    \label{fig:1}
\end{figure*}

\begin{figure*}[htbp]
    \centering
    \includegraphics[width=\linewidth]{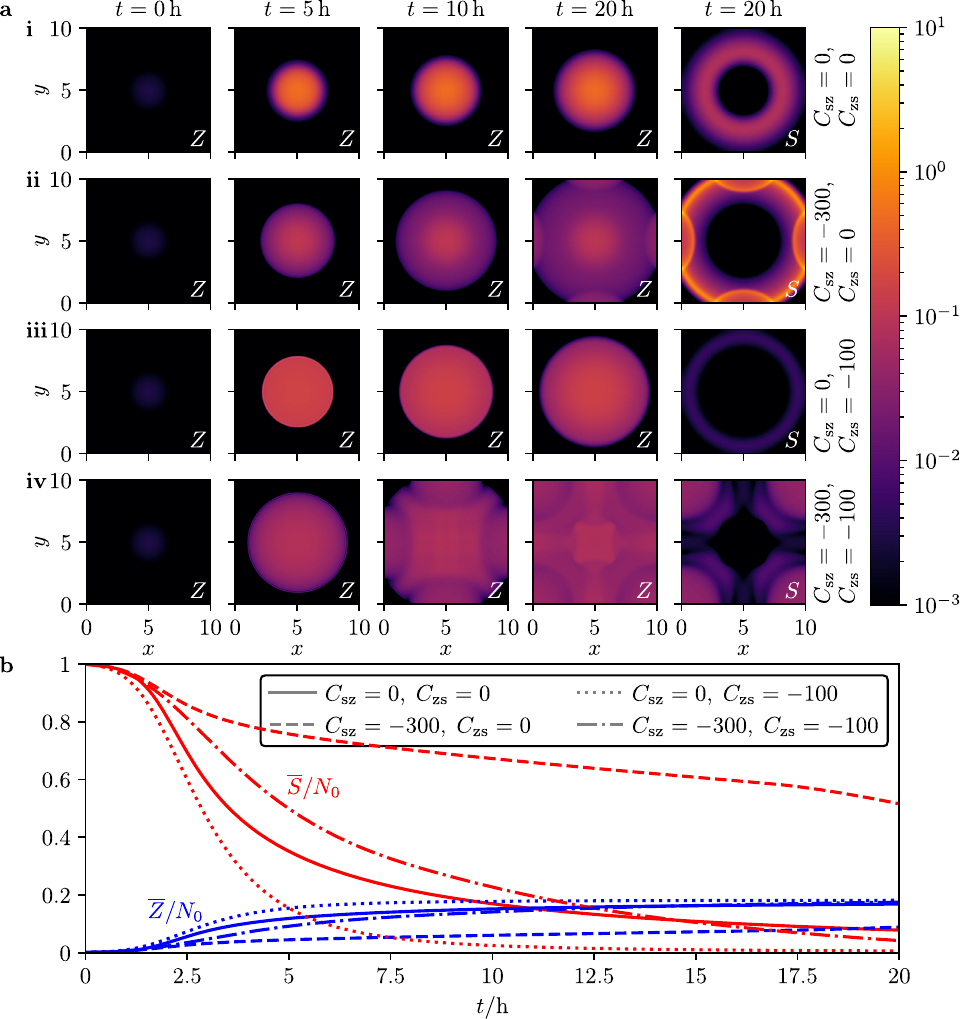}
    \caption{\textbf{(a)} Density of zombies $Z$ as a function of position $(x,y)^\mathrm{T}$ for times $t=0$ h, 5 h, 10 h, and 20 h, and density of susceptibles at $t=20$ h for interaction strengths (i) $C_{\mathrm{sz}}=C_{\mathrm{zs}}=0$, (ii) $C_{\mathrm{sz}}= - 300$, $C_{\mathrm{zs}}=0$, (iii) $C_{\mathrm{sz}}=0$, $C_{\mathrm{zs}}=-100$, and (iv) $C_{\mathrm{sz}}=-300$, $C_{\mathrm{zs}}=-100$. \textbf{(b)} Time evolution of the total number of susceptibles $\bar{S}$ and zombies $\bar{Z}$ for the four scenarios shown in (a).} 
    \label{fig:2}
\end{figure*}

\section{\label{simulations}Simulation of a zombie apocalypse}
To investigate the effect of non-reciprocal interactions in reaction-diffusion systems, we perform simulations of a zombie outbreak based on the SZ-DDFT model given by \cref{zombie1,zombie2}. Details on the numerical method are provided in Appendix \ref{numer}. We employ dimensionless units (except for time, which is measured in hours). 

\citet{WitkowskiB2013} have estimated the parameters of a zombie outbreak based on the popular zombie movie \textit{Shaun of the Dead} \cite{WrightP2004}. They found $b_{\mathrm{eff}} = 0.59$/h and $\kappa_{\mathrm{eff}} = 0.49$/h. In our case, the total population size is smaller than assumed in Ref.\ \cite{WitkowskiB2013} by a factor of about 10, which means that  $b_{\mathrm{eff}}$ and $\kappa_{\mathrm{eff}}$ should also be smaller by roughly this factor.\footnote{This can be seen from \cref{szrmodel1,szrmodel2,szrmodel3}. If $\bar{S}$ and $\bar{Z}$ are divided by a factor $N$, then \cref{szrmodel1} changes to $\dot{\bar{S}} = - b_{\mathrm{eff}}\bar{S}\bar{Z}/N$. To recover the original form, we have to absorb the factor $1/N$ into $b_{\mathrm{eff}}$.} We therefore use $b_{\mathrm{eff}}= 0.055$/h and $\kappa_{\mathrm{eff}}=0.045$/h. Assuming a homogeneous population and a domain size $A=100$, we can (in analogy to the argument employed for inferring the parameter $c$ in Ref.\ \cite{teVrugtBW2020b}) then get the parameters $b$ and $\kappa$ as $b=b_{\mathrm{eff}} A =5.5$/h and $\kappa = \kappa_{\mathrm{eff}} A$ = 4.5/h (assuming dimensionless units for length). The mean initial population density is $\hat{\rho}=0.25$, giving $N_0=\hat{\rho}A=25$.

Our goal is to investigate the optimal strategy of fighting a zombie outbreak. For this purpose, we perform a parameter scan in $\kappa$ and $C_{\mathrm{sz}}$ to generate a phase diagram, where we measure (a) the number of susceptibles at the end of the outbreak (number of survivors), (b) the number of zombies at the end of the outbreak, and (c) the time it takes for the battle of the living and dead to end. Thereby, we are able to compare two main strategies -- fighting the zombies, which corresponds to a large kill parameter $\kappa$, and running away from them, which corresponds to a large repulsive interaction strength $C_{\mathrm{sz}}$. We fix $b=5.5/\mathrm{h}$ (see above), $D_S=0.01$/h, $\Gamma_S=\Gamma_Z = 1$/h, and $\sigma_{\mathrm{sd}}=\sigma_{\mathrm{si}}=100$), and $D_Z = 0.005$/h (because zombies are slower than susceptibles). For the interaction strengths, we should note that the zombie disease is significantly more infectious than usual respiratory diseases (as seen from the large value of $b$ \cite{WitkowskiB2013}), such that the typical interaction strengths should be a factor 10 larger than in the case of the SIR-DDFT simulations in Ref.\ \cite{teVrugtBW2020} as zombies are very scary (and hungry). We therefore fix $C_{\mathrm{zs}} = -100$ and vary $C_{\mathrm{sz}}$ between 0 and $-300$. Similarly, we vary $\kappa$ between 0/h and 15/h to ensure that humans kill zombies at a rate whose order of magnitude is comparable to the rate at which zombies kill humans.

The resulting phase diagram is shown in \cref{fig:1}. We plot here, as a function of $C_{\mathrm{sz}}$ and $\kappa$, (a) the final number of susceptibles $\bar{S}_\infty$ in relation to the initial population size $N_0$, (b) the final number of zombies $\bar{Z}_\infty$ in relation to $N_0$, and (c) the time $t_{\mathrm{end}}$ that it takes for the zombie apocalypse to end. Notably, the parameter $C_{\mathrm{sz}}$ has no influence on the final number of susceptibles and zombies. All that matters is the kill rate $\kappa$. Consequently, for ending a zombie apocalypse, we have to kill the zombies and not run away from them. The determining factor is whether $\kappa$ is smaller or lager than 5.5/h, which is the value of $b$. For $\kappa < b$ all susceptibles are eliminated, whereas for $\kappa> b$, there is a large number of survivors. This number does not change when $\kappa$ is increased. Slightly more interesting is the dependence of $\bar{Z}_\infty$ on $\kappa$. While all zombies are eliminated for $\kappa > b$, the number of zombies that are still around after all susceptibles are eliminated depends on the kill rate. There is, for example, a large number of zombies at the end for $\kappa=0$, whereas for $\kappa$ slightly smaller than $b$, not only the number of remaining susceptibles, but also the number of zombies is very small. For $t_{\mathrm{end}}$, the overall picture is similar. The outbreak always ends very quickly for $\kappa > b$. In contrast, if $\kappa$ is increased from zero to a value below $b$, the time it takes till the battle is decided increases significantly (it even diverges for $\kappa = b$). Notably, there is also a small effect of increasing $|C_{\mathrm{sz}}|$ at constant $\kappa$ here, namely that it takes longer till all susceptibles are eliminated. Consequently, while running away does not end a zombie apocalypse, it does increase the time till the zombies bite everyone. 

In \cref{fig:2} a, the time evolution of the spatial distribution $Z(x,y,t)$ (with spatial coordinates $x$ and $y$) of the zombies is shown for some selected parameter values. For the final time $t=20$ h, we also show the distribution of the susceptibles $S$. In the noninteracting case ($C_{\mathrm{zs}} = 0$ and $C_{\mathrm{sz}} =0$, \cref{fig:2} a i), the zombies radially spread outwards, as in the SIR model with diffusion \cite{teVrugtBW2020}. The susceptible distribution at the final time looks like a ring. For $C_{\mathrm{zs}} = 0$, $C_{\mathrm{sz}} =-300$ (susceptibles are repelled by zombies, zombies are not affected by susceptibles, \cref{fig:2} a ii), the zombies move outwards radially up to a time $t=10$/h. Later, they are reflected at the boundaries of the system and move inwards, leading to a structure with four-fold symmetry. The distribution of susceptibles is still ring-like, but with a cross-shaped region that contains many susceptibles. If, on the other hand, the zombies are attracted by the susceptibles and the susceptibles simply move around randomly ($C_{\mathrm{zs}} = -100$, $C_{\mathrm{sz}} =0$, \cref{fig:2} a iii), the structures are generally similar to the ones observed in the noninteracting case, although there are fewer susceptibles at $t=20$/h.

The most interesting case is of course the one where both interactions are turned on ($C_{\mathrm{zs}} = -100$, $C_{\mathrm{sz}}=-300$, \cref{fig:2} a iv). Here, the zombies distribute  in space quicker, i.e., the field $Z$ spreads outwards faster than in the cases with no or fewer interactions. At the later stages ($t=10$ h), the model starts to show some interesting pattern formation that is different from what is known from the SIR-DDFT model with reciprocal interactions \cite{teVrugtBW2020}, where one finds concentric rings and later a separation into points that can be interpreted as infected persons self-isolating at their houses. In the present simulation based on the SZ-DDFT model, in contrast, one observes a square of zombies with bars at the edges on top of a spherical distribution at $t=10$ h, which then evolves into a smaller square with bars at the sides at $t=20$ h. The susceptibles, at this time, accumulate into quarter circles located at the edges.

It should be noted that the form of the observed structures, in particular their four-fold symmetry, is a consequence of the boundary conditions and the quadratic form of the simulation box. However, it is still interesting to show and discuss these structures here since (a) they differ quite significantly from what is observed in the reciprocal case, where the boundary conditions are the same but do not have a strong effect on the observed patterns, and (b) the boundary effects do have a physical relevance in the present context. If the susceptibles and zombies are in a quadratic domain, the susceptibles will accumulate at the edges in the final stages because this is the part of the domain that has not yet been conquered by the zombies. Thus, the simulation results show that for a zombie apocalypse, the spatial domain on which it takes place is considerably more important than in a normal pandemic. Pattern formation effects observed here are dominated by boundary effects, not by particle interactions as in the SIR-DDFT simulations performed in Ref.\ \cite{teVrugtBW2020}.

Figure \ref{fig:2} b shows the time evolution of the total number of susceptibles $\bar{S}$ and zombies $\bar{Z}$ in relation to the initial population size $N_0$ for the cases depicted in \cref{fig:2} a. In the noninteracting case with $C_{\mathrm{zs}} = 0$ and $C_{\mathrm{sz}} = 0$, the number of susceptibles declines relatively quickly. For $C_{\mathrm{zs}} = 0$ and $C_{\mathrm{sz}} =-300$ (susceptibles run away from the zombies and zombies  move around randomly), the number of susceptibles declines significantly slower, and there is a considerably larger number of survivors at $t= 20$/h. On the other hand, for $C_{\mathrm{zs}} = -100$ and $C_{\mathrm{sz}} = 0$ (zombies actively attack susceptibles and susceptibles do not actively run away), the zombies very quickly manage to kill essentially all the susceptibles. Finally, for $C_{\mathrm{zs}} = -100$ and $C_{\mathrm{sz}} =-300$ (susceptibles run away from zombies and zombies attack susceptibles), the overall number of survivors is very similar to the noninteracting case. Consequently, the two types of interactions compensate for each other (on the level of the entire population, the spatiotemporal dynamics is different), since at  $t= 20$/h one has a similar number of remaining susceptibles in the noninteracting and in the fully interacting case. Nevertheless, this final state is approached in a different way, with the initial decay of the number of susceptibles being slower than in the noninteracting case.

\section{\label{conc}Conclusions}
Being based on the SIR model, the SIR-DDFT model inherits the enormous flexibility of compartmental theories for epidemic spreading. In this work, we have demonstrated this flexibility by extending it towards vaccination, exposure and asymptomaticity, and mutations. We have also derived several extensions that are based on ideas from soft matter physics by incorporating noise and self-propulsion and by deriving an SIR-PFC model and a model with non-reciprocal interactions (describing a zombie outbreak). Of course, these extensions can also be combined, for example to generate a model that involves vaccination, mutations, and activity. Moreover, we have performed numerical simulations to study a zombie apocalypse, a scenario in which non-reciprocal interactions are relevant. In future work, our results can be used for modeling disease outbreaks in a more realistic way, and in particular to study chemical reactions using DDFT and PFC models in contexts where particle interactions (including non-reciprocal ones) and particle self-propulsion are relevant.

%**************************************************************************
%**************************************************************************

\acknowledgments{M.t.V.\ thanks the Studienstiftung des deutschen Volkes for financial support. R.W.\ is funded by the Deutsche Forschungsgemeinschaft (DFG, German Research Foundation) -- Project-ID 433682494 -- SFB 1459.}

\appendix

\section{\label{numer}Numerical methods}
The model equations \eqref{zombie1} and \eqref{zombie2} are solved by a finite difference scheme on a $512\times 512$
periodic grid in the case of \cref{fig:2} and a $256\times 256$ periodic grid in the case of \cref{fig:1}. The initial populations are given by Gaussian distributions centered at $x=L/2,y=L/2$ with the domain length $L$. The Gaussian has a variance of $L^{2}/75$ and is normalized such that the mean initial population density for the respective grid is equal to $\hat{\rho}=N_{0}/A=0.25$ and such that the ratio between the initial susceptible and zombie populations is given by $\bar{S}_{0}=999\bar{Z}_{0}$, i.e., on average one in every thousand persons is initially a zombie. For \cref{fig:2}, the time evolutions of
the fields $S$ and $Z$ were simulated for a total simulation time of $t=20$ h. For \cref{fig:1}, simulations were run until either the zombie or the susceptible population density fell below $5\cdot 10^{-4}\hat{\rho}$. From this, the final values $\bar{S}_{\infty}=\lim_{t\rightarrow\infty}\bar{S}$ and $\bar{Z}_{\infty}=\lim_{t\rightarrow\infty}\bar{Z}$ were estimated. In addition, we determined the time $t_{\mathrm{end}}$ at which either the total number of zombies or susceptibles fell below $5\%$ of the initial numbers $\bar{S}_{0}$ or $\bar{Z}_{0}$, respectively.

\bibliography{refs}
\end{document}